\documentstyle[12pt,aasms4]{article}

\def\msun{${\rm M}_\odot$}

\def\ps{s$^{-1}$}
\def\la{\mathrel{\hbox{\rlap{\hbox{\lower4pt\hbox{$\sim$}}}\hbox{$<$}}}}
\def\ga{\mathrel{\hbox{\rlap{\hbox{\lower4pt\hbox{$\sim$}}}\hbox{$>$}}}}

\begin{document}

\title{RXTE Observations of 0.1-300 Hz QPOs\\
in the Microquasar GRO J1655-40}
\author{Ronald A. Remillard, Edward H. Morgan}
\affil{Center for Space Research, Massachusetts Institute of Technology, Room 37-595, Cambridge MA 02139}
\authoremail{rr@space.mit.edu.edu; ehm@space.mit.edu}
\author{Jeffrey E. McClintock}
\affil{Harvard-Smithsonian Center for Astrophysics, 60 Garden St., Cambridge MA 02138}
\authoremail{jem@cfa.harvard.edu}
\author{Charles D. Bailyn}
\affil{Department of Astronomy, Yale University, P. O. Box 208101, New Haven, CT 06520}
\authoremail{bailyn@astro.yale.edu}
\and
\author{Jerome A. Orosz}
\affil{Dept. of Astronomy and Astrophysics, Pennsylvania State University, 525 Davey Laboratory, University Park, PA 16802}
\authoremail{orosz@astro.psu.edu}

\begin{abstract}
We have investigated 52 RXTE pointed observations of GRO J1655-40
spanning the X-ray outburst that commenced on 1996 April 25 and lasted
for 16 months.  Our X-ray timing analyses reveal four types of QPOs:
three with relatively stable central frequencies at 300 Hz, 9 Hz, and
0.1 Hz, and a fourth that varied over the range 14--28 Hz.  The 300 Hz
and 0.1 Hz QPOs appear only at the highest observed luminosities ($L_X
> 0.15 L_{Edd}$), where the power-law component dominates the X-ray
spectrum, and the inner disk is characterized by a $\sim 1.5$ keV
color temperature and an estimated inner radius of only 10-20 km.  At
lower luminosity ($L_X \sim 0.1 L_{Edd}$), the thermal component
dominates the spectrum; the disk is somewhat cooler ($\sim 1.3$ keV)
and its inner radius is larger. In this state only two of the QPOs are
observed: the broad and spectrally "soft" 9 Hz QPO, and the narrow,
"hard" QPO that varies from 14-28 Hz as the hard flux decreases.  At
still lower luminosities ($L_X < 0.1 L_{Edd}$), the power-law
component contributes less than 30\% of the total luminosity, the
inner disk appears both larger and cooler, the 9 Hz QPO vanishes, and
only a very weak (rms 0.3\%) and narrow QPO at 28 Hz remains.

The 300 Hz QPO is likely to be analogous to the stationary QPO at 67
Hz seen in the microquasar GRS1915+105.  We discuss four models of
these high-frequency QPOs which depend on effects due to general
relativity.  The models suggest that these rapid QPOs may eventually
provide a measure of the mass and rotation of the accreting black
hole. The 9 Hz QPO displays a spectrum consistent with a thermal
origin, but this frequency does not appear to be consistent with any
of the natural time scales associated with the disk, or with the
inferred values of the mass and rapid spin of the black hole.  The
mechanism for the 14-28 Hz QPOs appears to be linked to the power-law
component, as do the 1-10 Hz QPOs in GRS1915+105. Thus, these
low-frequency QPOs have the potential to lead us to the origin of the
energetic electrons that radiate the power-law spectral component.
Finally, we show data for GRO J1655-40 and GRS1915+105 as each source
teeters between relative stability and a state of intense oscillations
at 0.1 Hz.  A comparison of the sources' spectral parameters allows us
to speculate that the black hole mass in GRS1915+105 is very large --
possibly in the range 39-70 \msun.

\end{abstract}


\section{Introduction}

The black hole binary GRO J1655-40 is one of the rare Galactic X-ray
sources (along with GRS1915+105 and possibly Cyg X-3) known to produce
relativistic radio jets (\cite{mir94}; \cite{tin95}; \cite{hje95};
\cite{mio97}; \cite{new98}). These jet sources, and a few other
Galactic X-ray sources that exhibit double-lobed radio structure
(e.g. 1E1740.7-2942; see \cite{smi97} and references therein), are
collectively referred to as ``microquasars'', since they have
properties analogous to radio-loud active galactic nuclei.
Investigations of microquasars provide opportunities to search for the
cause of relativistic jets, to probe the properties of accreting black
holes and their mass-donor companion stars, and to measure the
spectral evolution associated with the various accretion states.  GRO
J1655-40 is an especially important target for such studies, since its
distance and black hole mass are well established: $d \sim$ 3.2 kpc
has been inferred from radio absorption features (\cite
{tin95}; \cite {mck94}), and $M_1 \sim 7.0$ \msun\ has been derived from
optical investigations of the binary system (\cite {bai95}; \cite {orb97}).

GRO J1655-40, was discovered with BATSE on 1994 July 27. For about
four months thereafter a correlation between hard X-ray outbursts
(20--100 keV) and radio flares was observed, and this established a
relationship between jet formation and hard X-ray activity
(\cite{har95}). However, subsequently during 1994--95 the hard X-ray
outbursts continued unabated while the radio responses decreased
(\cite{tav96}). Thus jet formation in GRO J1655-40 depends on some
unknown factor(s) in addition to hard X-ray outbursts.

The X-ray spectrum of GRO J1655-40 has been detected with OSSE to
photon energies above 1 MeV (\cite{kro96}; \cite{tom98}). Above 50
keV, the spectrum is consistent with a power law function and a photon
index in the range of 2.4--2.8.  Soft X-ray observations were made
with ROSAT during 1994 August (\cite {gre95}) and with ASCA during
1994 August--September and 1995 August (\cite {ued98}).  The detected
flux (2--10 keV) was in the range of 0.2--4.7$\times 10^{-8}$ erg
cm$^{-2}$ \ps\, with a suggestion of persistence in soft X-rays during
the times between hard X-ray outbursts.  Below 10 keV, the spectrum is
dominated by a component which is attributed to thermal emission from
the inner accretion disk. 

Spectral fits for this thermal X-ray component indicate that the size
of the inner disk in GRO J1655-40 appears to be relatively small and
hot, compared to the disks of other black hole binaries.  A
nonrotating 7 \msun\ black hole would be expected to have an inner disk
larger than the innermost stable orbit of 6 $GM_1/c^2$ or 62 km, while
a rapidly rotating (prograde) black hole may have a last stable orbit
as small as 1 $GM_1/c^2$ or 10 km (see \cite{zha97} and references
therein). GRO J1655-40 appears to have an inner disk radius of 22 km
or 2.1 $ GM_1/c^2$ (\cite{zha97}), which was interpreted as evidence
that the black hole is rotating at $\sim$93\% of the maximum rate
(\cite{zha97}). Such rapid rotation may be the distinguishing
characteristic of the microquasars.

During late 1995 and early 1996, GRO J1655-40 entered a very low or
quiescent accretion state (e.g. \cite{ued98}), permitting a clear
optical view of the companion star. This allowed Orosz \& Bailyn (1997)
to improve the determinations of the binary period (2.62 d), the mass
function (3.2 \msun), and the spectral type of the secondary (F4IV).
They further modeled the ``ellipsoidal variations'' of the secondary
star and partial eclipses of the accretion disk. Their analysis, using
B,V,R, and I bandpasses, provides an exceptionally good fit for the
binary inclination angle ($69.5^{\circ}$) and the mass ratio.  From these
results, they deduce masses of $7.0 \pm 0.2$ and $2.34 \pm 0.12$ \msun\
for the black hole and companion star, respectively (also see
\cite{vdh98}). This is the only microquasar for which there is a clear
understanding of the masses and dimensions of the binary system.

The All Sky Monitor (ASM) aboard the {\it{Rossi X-ray Timing
Explorer}} (RXTE) began regular monitoring of the X-ray sky at 2--12
keV in early 1996. For the first two months of ASM observations, the
source was quiescent and not detected. Then, on 1996 April 25 a new
outburst was discovered (\cite{rem96}). With great fortune, concurrent
optical monitoring was in progress, and it was determined that optical
brightening preceded the X-ray turn-on by 6 days, beginning first in
the I band and then occurring sequentially in the R, V, and B bands
(\cite{oro97}). These results provide evidence favoring a
two-zone accretion flow in quiescent X-ray novae: a cold outer
disk and an extremely hot inner disk with an advection-dominated
accretion flow (\cite{ham97}).

Herein we report the results of X-ray timing investigations derived
from a series of 52 RXTE pointed observations of GRO J1655-40 covering
the 1996--1997 outburst. A brief and preliminary account of these
results has been reported previously (\cite {rem97}). We describe four
types of transient quasi-periodic oscillations (QPOs) in X-rays that
span 0.1 to 300 Hz.  Three of these QPO types appear only when the
luminosity of the hard power-law component is high. Moreover, the 300
Hz QPO is seen only when the power-law component dominates the
spectrum, and the total X-ray luminosity exceeds $\sim$10--15\% of the
Eddingtom limit ($L_{Edd}$). This fast QPO appears to be analogous to
the stationary 67 Hz QPO seen in GRS1915+105 (\cite {mor97}; \cite
{rem98}).

\section{Observations with the RXTE All Sky Monitor}

The RXTE ASM scans the celestial sphere roughly five times per
day. Each of its three cameras contains a position-sensitive
proportional counter that is mounted below a coded mask. The typical
sensitivity for measurements averaged over one day is $\sim 7$ mCrab
at 3 $\sigma$ significance over the full energy range of the
instrument. Data in the form of position histograms are accumulated in
three energy channels: 1.5--3.0, 3--5, and 5--12 keV. The mask shadow
patterns are deconvolved to yield X-ray source intensities, which are
then normalized to the response characteristics of camera \#1. Further
details regarding the instrument and data analysis methods are
described by \cite{lev96} (1996). 

The ASM light curve for GRO J1655-40 is shown in Figure
\ref{fig:asm}. These data have been renormalized via the instrument
calibrations that were revised on 1998 March 15. The top panel
displays the X-ray count rate at 2--12 keV; the vertical lines above
the data points indicate the times of RXTE pointed observations which
are of central importance to this paper. As shown below in Section 4,
the hard X-ray component is usually sufficiently steep that the 2--12
keV count rate is a fairly good representation of the overall X-ray
luminosity. The bottom panel of Figure~\ref{fig:asm} shows
measurements of the spectral hardness ratio, {\it{HR2}}, which is defined
as the ASM count rate at 5--12 keV relative to the rate at 3--5
keV. The overall profile of this outburst is markedly different from
the fast rise and slow exponential decay characteristic of many X-ray
novae (\cite{che97}).  In the case of GRO J1655-40, the first wave in
X-ray brightness lasts for about 270 days. Initially the source is in
a soft state, but then it evolves into a hard, flaring state. The
highest flux during these flares (285 c/s) corresponds to an X-ray
luminosity (unabsorbed) near $2\times10^{38}$ erg s$^{-1}$ at 2--12
keV, which is $\sim$0.2 $L_{Edd}$ for GRO J1655-40. The isolated low
points in the ASM light curve have been shown to be associated with
absorption dips that occur in a narrow range of binary phase (\cite
{kuu98}).

After MJD 50400, the X-ray flux decreases while the spectrum softens,
reaching a local minimum near MJD 50460. However, this evolution is
interrupted by a second wave in brightness that extends the outburst
for another 200 days. The second maximum also displays an interval of
increased variability and spectral hardening.  However, compared to
the first maximum, the intensity is substantially lower and the
flaring is subdued. The decay from the second maximum is again
associated with spectral softening.  However beginning at MJD 50670
(1997 August 10), at a count rate of 12 c/s (or only 0.008 $L_{Edd}$),
the spectrum abruptly hardens again.  The flux continues to decay, and
on MJD 50681 (1997 August 21) the source falls below the ASM detection
limit (3$\sigma$ upper limits $\sim$9 mCrab per 1-day bin).
The ASM results provide a thorough context for
the analysis and interpretation of the RXTE pointed observations reported
below. Beyond the results shown in Figure~\ref{fig:asm}, there have
been no further ASM detections of GRO J1655-40 through the end of 1998
May (MJD 50965).

\section{PCA Power Spectra of GRO J1655-40}

The central theme of this paper is an X-ray timing analysis of a
series of 52 observations with the RXTE Proportional Counter Array
(PCA). The instrument is described by \cite{jah96} (1996). The PCA
consists of five xenon-filled detectors with a total collecting area
$\sim$ 6200 cm$^2$ at 5 keV, sensitivity in the range of 2--60 keV,
and $\sim17 \%$ energy resolution at 5 keV. Depending on the observing
modes chosen for a particular observation, the time resolution may be
as fine as 1 $\mu$s.

The PCA observation times are listed in Table~\ref{tab:obs}; they are
distributed throughout the 1996--1997 outburst, as shown in Figure
~\ref{fig:asm}.  Table~\ref{tab:obs} also provides exposure times, the
number of detector units (PCUs) that were collecting data, and the PCA
observing modes. The data in this study include all of the
observations in our AO-1 guest observer program (P10255), and also the
observations for the AO-2 public archive (P20402), except for those
data obtained after MJD 50685 (1997 August 25) when the source
intensity was below 1 mCrab. With one exception, we define an
observation as the collection of PCA data obtained on any one
particular UT day. On 1996 November 2, GRO J1655-40 changed its intensity 
and QPO properties significantly between intervals of good exposure
time, and we therefore divided these data into two observations labeled
``A'' and ``B'', respectively (see Table~\ref{tab:obs}).

In Table~\ref{tab:results} we provide a variety of measurements that
describe the X-ray emission properties of GRO J1655-40. These include
the PCA count rates for three energy bands (cols. 3, 7, and 8), the
standard deviation of the full (effectively 2--25 keV) PCA count rate
in 1 s bins (col. 4), two types of PCA hardness ratio (cols. 5, 6),
and QPO frequencies and comments relevant to the observations
(col. 9). These results will be used below to help evaluate the
conditions under which the different types of QPOs appear.

\subsection{Power Spectra for Individual Observations}

In Figures~\ref{fig:pds1}--\ref{fig:pds3} we show the power spectra
for each of the 52 PCA observations of GRO J1655-40 listed in
Table~\ref{tab:obs}. The corrections for instrument dead time and
electrical recovery from very large events adopts the model for
``paralyzable'' effects described by Morgan, Remillard, \& Greiner
(1997).  In fact, the instrument operates with a combination of
paralyzable and non-paralyzable interruptions (\cite{zha95}), and our
corrections leave behind residual broadband power at the level of
0.1\% Hz$^{-1}$ of the net count rate. Since our power spectra are
normalized to represent the square of the fractional rms amplitude per
Hz, any broad continuum at high frequencies in Figures
\ref{fig:pds1}--\ref{fig:pds3} with a power density $\la 10^{-6}$
Hz$^{-1}$ is likely to represent residual instrumental effects rather
than a real signature of source behavior.

All of the power spectra in Figures~\ref{fig:pds1}--\ref{fig:pds3}
show a broadband continuum that typically decreases with frequency in
a manner roughly described as a power law with an index of 1.2 ($\pm
0.3$).  Frequently there is very broad curvature in this continuum in
the log power vs. log frequency plane, extending from 0.1--10 Hz with
an excess centered near 1 Hz.  During many observations, we also see
QPOs in the range of 7--28 Hz. The profiles of these QPOs are
complex. In several cases (e.g. 07/25/96, 08/16/96, 08/22/96,
09/04/96, 09/09/96, 10/27/96), the profiles suggest separate broad and
narrow QPOs, with the latter found at higher frequency.
Furthermore, there are two occasions (8/01/96 and 11/02/96B) in which
there is an additional low-frequency QPO near 0.1 Hz.

During the final decay of GRO J1655-40 in 1997 August (after MJD
50663), the power continuum abruptly increases and flattens, and there
are renewed QPOs with relatively narrow profiles. These changes in the
power spectra coincide with the episode of spectral hardening seen in
the ASM (see Fig.~\ref{fig:asm}). Further spectral analysis of this
low-hard state is deferred to a later publication.

To derive the central frequencies and widths of each QPO, we fit the
power spectra to a local continuum, using a power law function, plus a
QPO assumed to have a Lorentzian profile.  We used a least-squares
technique to determine the best fit. In the range of 7--28 Hz, we
allowed for the presence of two overlapping QPOs with different
widths. The locations of the continuum fitting regions had to be
adjusted for the particular QPO locations and broadband power
characteristics of each observation.  The central QPO frequencies
derived from these fits are included in Table~\ref{tab:results}, while
all of the QPO parameters for individual detections in the range of
0.1--28 Hz are given in Table~\ref{tab:qpos}. These parameters include
the central frequency ($\nu$), the FWHM value (in Hz), the coherence
parameter ($Q = \nu / $FWHM), and the QPO amplitude, which is the
integrated power normalized to the mean count rate of GRO J1655-40 in
the appropriate energy band.

The data summarized in Table~\ref{tab:results} show that the intensity
and spectrum of the source can be used to predict the presence of QPOs
at 7--28 Hz.  Excluding the low/hard state at the end of the 1996--97
outburst, the approximate thresholds for QPO activity are: a 2--25 keV
count rate of 4000 c \ps\ pcu$^{-1}$, a 9--25 keV count rate of 200 c
\ps\ pcu$^{-1}$, a value of 0.45 in {\it{PCA HR1}} (counts at 5-9 keV
/ counts at 2-5 keV), or a value of 0.14 in {\it{PCA HR2}} (counts at
9-13 keV / counts at 5-9 keV). These thresholds are all related to the
appearance of substantial flux above 9 keV, which suggests that the
QPO mechanism is linked to the hard power-law component and thus to
the source of relativistic electrons. In the case of GRS1915+105, the
increase in QPO amplitudes with photon energy supported a similar
conclusion (\cite{mor97}).  Further considerations regarding the QPO
mechanism are given in Section 5.

\subsection{Detection of a QPO at 300 Hz}

The relationship between QPO activity and the hard X-ray spectrum
motivated us to combine the individual power spectra in groups
determined by their spectral hardness, in order to search for more
subtle features, especially at higher frequency. Using the {\it{PCA
HR1}} values given in Table~\ref{tab:results}, we sorted and combined
the individual power spectra (excluding observations 50--52) into four
groups: 1996 and {\it{HR1}} $< 0.45$, 1997 and {\it{HR1}} $< 0.45$,
$0.45 <$ {\it{HR1}} $< 0.50$, and {\it{HR1}} $> 0.50$. We label these
groups, respectively, ``soft 1996'', ``soft 1997'', ``hard'', and
``very hard''.  For each group the individual power spectra (2--25
keV) were averaged after weighting them by their respective exposure
times. The results are shown in Figure~\ref{fig:pdssum}, where four
important details can be noted. (1) The 27 power spectra in the group
``soft 1997'' do provide a narrow QPO just below 30 Hz. (2) There is a
suggestion that the QPOs move to lower frequency as the spectrum
becomes harder. (3) The QPO near 0.1 Hz is apparent in the ``very
hard'' group. (4) Finally, and most importantly, there is a broad bump
near 300 Hz for the ``very hard'' group.

In Figure~\ref{fig:pdssumz} we replot Figure~\ref{fig:pdssum} with a
magnified scale to provide closer inspection of those QPO features
above 10 Hz. We fit the profiles for two of the QPOs, as described
above.  The QPO for the ``soft 1997'' group is located at 28.3 Hz with
a FWHM of 3.6 Hz and an integrated amplitude of 0.31\% over the
effective PCA bandwidth (2--25 keV). In the remainder of Section 3.2
we focus solely on the broad feature near 300 Hz.

The QPO fit at 300 Hz for the ``very hard'' group
(Figure~\ref{fig:pdssumz}) yields a central frequency of 298.3 Hz, a
FWHM of 77 Hz, and an integrated amplitude of 0.45\%. The significance
of this QPO detection above the local power continuum is 6.2 $\sigma$
in the highest bin and 12 $\sigma$ as an integrated feature .

The origin of the 300 Hz feature within the ``very hard'' group is
investigated in Figure~\ref{fig:pdsstack}.  Here the A and B
observations of 1996 Nov 2 have been combined. The fit for the QPO
profile and the local power continuum is shown with a solid line.  A
weak power excess near 300 Hz is seen in most of the individual power
spectra, especially in the five power spectra for 1996 August. We
conclude that the 300 Hz feature is statistically diluted among the
individual power spectra, as would be expected for observations of a
very weak yet stationary process.

If we select only the 1996 August observations within the ``very
hard'' group (i.e. the top 5 panels of Fig.~\ref{fig:pdsstack}) and
refit the continuum and QPO profile, then the detection significance
of the integrated 300 Hz feature is again 12 $\sigma$, and the other
results are changed only slightly: the central frequency is 297.5 Hz,
the FWHM is 68 Hz, and the integrated amplitude is 0.48\%.

The upper limits (3 $\sigma$) for a QPO in the range of 250--350 Hz
(with FWHM of 68 Hz) are 0.21\% of the mean flux for the ``hard''
group, 0.16\% for the ``soft 1997'' group, and 0.32\% for the ``soft
1996'' group. We therefore conclude that the 298 Hz QPO in GRO
J1655-40 is significantly weaker or absent when the luminosity in the
power-law component falls below that of the ``very hard'' group
(i.e. when {\it{HR1}} $< 0.5$).

Finally, the effort to ascertain whether the frequency of this QPO is
constant is limited to the weak indications provided in
Figure~\ref{fig:pdsstack}.  We searched for a QPO with a central
frequency in the range of 180--420 Hz for each individual panel shown
in Figure~\ref{fig:pdsstack}; the power spectra in the top six panels
yield a sample of central frequencies that are distributed as 300.2
$\pm$ 13.4 Hz (see Table~\ref{tab:results}). The variations in
frequency are consistent with scatter due to statistical
limitations. Meanwhile, the count rates in the hard PCA band
(i.e. above 9 keV) for these same six observations vary by a factor of
2.5 (max / min). Since the flux in hard photons is highly correlated
with the appearance of this QPO, and since the hard flux variations
are much larger than the dispersion in central frequency, we regard
these results as an indication of frequency stability in the 300 Hz
feature.

\subsection{QPO Properties versus Photon Energy}

Our detection of the 300 Hz QPO in GRO J1655-40 is limited to just one
energy channel (2--25 keV) for the following reason. The operating
modes for the PCA event analyzers (EAs) are given in
Table~\ref{tab:obs}.  During the ``very hard'' observations (1996
August - November), the four discretionary PCA EAs were operated in
parallel in the following modes: (1) a 4-channel binned mode with 2 ms
time resolution at 2--13 keV; (2) a single-channel, ``single bit''
mode with 62 $\mu$s time resolution at 2--13 keV; and (3) a 16 $\mu$s
event mode for photons above 13 keV. The Nyquist frequency for 298 Hz
oscillations corresponds to 1.7 ms, which is faster than the rate used
in the 4-channel binned mode. We therefore have only two channels,
divided at 13 keV, with which to examine phenomena at frequencies
above 250 Hz. During the ``very hard'' observations, the 13--25 keV
source count rate was only 10--18 \% of the rate below 13 keV, and we
do not detect a significant QPO in these data.  The PCA time
resolution was increased to 4-channel coverage with 125 $\mu$s time
resolution for the 1997 observations, but there were no further
episodes of ``very hard'' emission during this period.

For the low frequency QPOs, however, we can easily measure the QPO
properties for ``hard'' and ``very hard'' observations in five
different energy bands, with four bands telemetered in binned mode and
the highest energy band obtained from event mode. We show two
energy-resolved series of power spectra in Figure~\ref{fig:pdsen}. The
left panels show results for an observation in the ``very hard'' group
(08/16/96), while the right panels show results for an observation in
the ``hard'' group (09/09/96) . Both cases show a broad QPO near 8--10
Hz combined with a narrower QPO at higher frequency. The central
frequencies for the 2-QPO fits (2--25 keV) for each observation are
shown with arrows in the top and bottom panels. The energy-resolved
power spectra in Figure~\ref{fig:pdsen} show that the broad and narrow
QPOs have very different energy dependence: in both observations, the
broad QPO is stronger at lower photon energies, while the narrow QPO
shows an increase in amplitude at higher photon energies.

Further insight into the broad-soft and narrow-hard QPOs are gained by
plotting the central frequencies for the 2-QPO fits versus the PCA
count rate in the hard band (9--25 keV). The results are shown in
Figure~\ref{fig:fxqpo}. The plotting symbols distinguish broad and
narrow QPOs using the coherence parameter, $Q$, as defined previously.
The ``*'' symbol indicates broad QPOs with $Q < 3.0$, while the open
triangles denote narrow QPOs with $Q > 3.0$. The data point marked
``x'' represents the narrow and weak QPO for the entire ``soft 1997''
group, as shown in Figure~\ref{fig:pdssumz}. This 28 Hz QPO appears to
be a simple extrapolation of the narrow QPO branch shown in
Figure~\ref{fig:fxqpo}. The frequency of narrow QPOs is correlated
with the intensity in hard X-rays, shifting from 28 to 14 Hz as the
hard intensity increases from zero to maximum.  This is the only QPO
that is visible as the count rate at 9--25 keV approaches zero.  On
the other hand, the broad QPOs near 9 Hz in Figure~\ref{fig:fxqpo}
remain essentially stationary, and this QPO is not observed when the
soft component dominates the spectrum. There are some exceptions to
the pattern of QPO branches in Figure~\ref{fig:fxqpo}. For example, we
do not detect two distinct QPOs for every observation in the hard
groups, and two of the QPO detections near 9 Hz appear to be
relatively narrow in FWHM.  Nevertheless, the weight of the evidence
suggests that there are two quite different and yet coexisting types
of QPOs in GRO J1655-40 in the range of 7--28 Hz.

\subsection{QPOs at 0.1 Hz and a Connection to GRS 1915+105}

The power spectra of 08/01/96 (Fig. 2) and 11/02/96B (Fig. 3) each
display a low-frequency QPO near 0.1 Hz. These same two observations
are at the extreme end of the ``very hard'' group, yielding the highest
values for the X-ray intensity at 9--25 keV, as shown in
Table~\ref{tab:results} and Figure~\ref{fig:fxqpo}.  The 0.1 Hz QPOs lie
atop substantial broadband power, and their integrated rms power
values (2.9 and 4.4\% of the mean count rate, respectively) rank them
among the QPOs in GRO J1655-40 with the highest amplitudes (see
Table~\ref{tab:qpos}).

The appearance of low frequency QPOs during conditions of high X-ray
luminosity in GRO J1655-40 is reminiscent of the behavior of the
microquasar GRS1915+105. The latter source is prone to episodes of
tremendous X-ray variability (\cite{gre96}), but there are also
occasions when the X-ray intensity of GRS1915+105 is relatively low
and steady (with 1-10 Hz QPOs), and there are other occasions
characterized by intermediate brightness, moderate flickering and QPOs
near 0.1 Hz (\cite{mor97}). In particular, the power spectra of GRS
1915+105 on 1996 April 17--May 14, 1996 September 22, 1997 August 13,
and 1997 November 7--30 show QPOs near 0.1 Hz. The corresponding light
curves show substantial flickering but there are no large cyclical
variations and the rms deviation (1 s bins) is less than 30\% of the 
mean value, which is also true of our light curves of GRO J1655-40
(excluding absorption dips).

There is an especially striking resemblance between the PCA
observation of GRO J1655-40 on 1996 August 1 and the observation of
GRS 1915+105 on 1997 August 13. In each case, the source appears to
shift back and forth between a relatively steady emission state and
another state in which both the intensity and the hardness ratio are
strongly modulated by a 0.1 Hz QPO. This behavior is illustrated for
both sources in Figure~\ref{fig:lcflick}. Both the X-ray intensity and
variability of GRO J1655-40 are as large as we have seen with
RXTE. However, GRS1915+105 has shown a wide variety of much deeper
oscillations on time scales ranging from seconds to $\sim$30 min.
Given the similarities shown here and the extraordinary behavior
observed for GRS 1915+105 (\cite{gre96}), one may surmise that GRO
J1655-40 is on the verge of plunging into unstable X-ray emission on
1996 August 1.  Figure~\ref{fig:lcflick} displays the closest link we
have seen between GRO J1655-40 during its radio-quiet X-ray outburst
of 1996-1997, and GRS 1915+105, which has been wildly active at both
X-ray and radio frequencies (\cite{eik98}; \cite{mir98};
\cite{poo98}). In Section 5 we use these intermittent 0.1 Hz QPOs,
combined with spectral analyses and our knowledge of the black hole
mass in GRO J1655-40, to speculate on the mass of the black hole in
GRS1915+105.

\section{X-ray Spectral Analysis by QPO Group}

Thus far we have described several relationships between QPO
characteristics and the X-ray spectrum by correlating the QPO
parameters with PCA count rates and hardness ratios. In these
interpretations we have assumed that the X-ray spectrum of GRO
J1655-40 can be described by a standard model consisting of a thermal
component and a hard power law, with a minimal contribution from the
thermal component above 9 keV.  Zhang et al (1997) used ASCA and BATSE
observations of GRO J1655-40 during 1995 August 15-16 to show that
this spectral model produced an acceptable fit with an inner disk
temperature of 1.36 keV. A spectral analysis of all of the data
discussed herein is beyond the scope of this paper. Nevertheless, we
have investigated the mean spectra for each QPO-sorted group (see
Section 3.2) and the results confirm our hypothesis that QPO activity
in GRO J1655-40 is strongly linked to the strength of the power-law
component.

We used ``xspec'' to analyze these data, assuming a ``disk blackbody''
model plus a power-law function, both attenuated by absorption due to
cold gas along the line of sight. The analysis was performed with
version 2.2.1 of the PCA spectral response matrices (written on 1997
Oct 2). Since the high count rates and long exposure times for GRO
J1655-40 stringently test the quality of the PCA spectral
calibrations, we first investigated the spectra derived from long
exposures of the Crab nebula on 1997 March 22, April 15, and July 26.
We have concluded that the best performance from these particular
response matrices is gained by choosing PCUs \#0, \#1, and \#4 (only),
by restricting the analysis range to 2.5--25.0 keV (with 54 data
channels from PCA standard mode 2), and by imposing a systematic error
of 1.0\% on each data point. The statistical error bars for the
spectral parameters in each group are very small, and so we
conservatively choose to independently analyze the spectra provided by
each of the three PCUs, and then we use the sample standard deviation
as the uncertainty in each spectral parameter. The absolute
normalizations of PCU \#4 are chronically 10\% lower than those of
PCUs \#0 and \#1, and we accordingly raised the normalizations for PCU
\#4 before reporting the results. Finally, all of the spectral results
(unlike the counting rates reported in Table 2) include corrections
for the PCA deadtime, which ranges from 1.5\% to 12.8\% per
observation, depending on the source count rate.

We conducted spectral fits in which we considered the value of the
interstellar column density ($N_H$) as a free parameter, and then we
repeated the analysis with $N_H$ = 0.9$\times 10^{22}$ cm$^{-2}$,
which is the value found with ASCA (\cite{zha97b}). The results for
both cases are shown in Table~\ref{tab:spec}. Although the fitted
versus fixed values of $N_H$ are statistically different in most
cases, this difference has only a minor effect on the other spectral
parameters or the integrated flux.

The reduced $\chi^2$ values (Table~\ref{tab:spec}, col. 4) for these
spectral fits are reasonably good for the ``hard'' and ``very hard''
groups, with total exposure times $\sim$40 ks per PCU for each
group. On the other hand, the ``soft 1996'' group resisted all efforts
to apply the disk blackbody plus power law model, including attempted
modifications for emission lines, absorption edges, or Compton
reflection. Separating the ``soft 1996'' group into subsections A
(1996 May 9--12) and B (Sept. 26 and Oct. 3) isolates the temporal
location of this unsolved problem.  During the 1996 May observations,
which occurred 14-17 days after the beginning of the X-ray outburst,
we find $\chi_{\nu}^2\sim20$, as the soft component appears to be too
wide to fit with any combination of disk ``color temperature''
($T_{col}$) and $N_H$. No such difficulty is evident for the ``soft
1996B'' group ($\chi_{\nu}^2 = 1.04$).  Problems with the spectral
model are encountered again for the ``soft 1997'' group.  However, the
residuals are only $\sim2$\% of the net flux in the accumulated
spectra (102 ks per PCU), and we argue that the comparison of spectral
parameters with the other groups provides useful information, despite
the elevated value of $\chi_{\nu}^2$. As the overall luminosity
increases, it is clear that the temperature of the inner accretion
disk increases from $\sim$1.1 to nearly 1.5 keV, while the
normalization decreases by a factor of 5. This is the same pattern
seen on a much more rapid time scale in the case of GRS1915+105
(\cite{bel97b}).  Further evaluations of these results are given in
Section 5.

Table~\ref{tab:spec} confirms that the flux in the power-law component
rises sharply as the spectrum evolves from the ``soft'' group to
the ``hard'' group, and on to the ``very hard'' group, as we had
deduced from the PCA hardness ratios. In the ``hard'' group, the
power-law component represents $\sim$30\% of the total X-ray flux, and
this fraction increases to $\sim$70\% in the ``very hard'' group. In both
the ``hard'' and ``very hard'' groups, the photon index is in the
range of 2.4--2.7, which is very similar to the results obtained in
earlier outbursts for the range of 20--600 keV with CGRO
(\cite{kro96}; \cite{zha97b}).  Our 1996 August 29 PCA observation was
used with RXTE HEXTE and CGRO OSSE observations to measure a photon index
of 2.7 with a lower limit of 800 keV for the cutoff energy associated
with the power law component (\cite{tom98}). Our similar value of the
photon index for the ``very hard'' group instills some confidence that
the power law component is well determined here, despite the constraint
which limits our spectral fits to photon energies below 25 keV.

Our spectral results allow us to quantify the thresholds for QPO
activity in terms of $L_{Edd}$, which is 9 $\times 10^{38}$ erg \ps\
for a 7 \msun\ black hole. The threshold where the broad 9 Hz QPO
appears is $L_X \sim 0.1 L_{Edd}$. We detect 300 Hz QPOs at $L_X \ga
0.15 L_{Edd}$, and 0.1 Hz QPOs appear at a slightly higher value,
$L_X \sim 0.18 L_{Edd}$. For these values to have utility, the
power-law component must radiate isotropically from a region that is
roughly as compact as the inner accretion disk.

\section{Discussion}

The 300 Hz QPO in GRO J1655-40 is the fastest oscillation ever seen in
a black hole binary.  Efforts to explain this feature, and also the 67
Hz QPO in GRS1915+105, commonly invoke effects of General Relativity
(GR), and these results provide an important opportunity to
investigate both the properties of these black holes and also the
physics of strong gravitational fields (see \cite{mcc98}). There are
at least four proposed mechanisms that relate these fast QPO
frequencies to a natural time scale of the inner accretion disk in a
black hole binary.  These are: the last stable orbit (\cite{sha83};
\cite{mor97}), diskoseismic oscillations (\cite{per97}; \cite{now97}),
frame dragging (\cite{cui98}), and oscillations related to a
centrifugal-barrier model (\cite{tit98}). The physics of all of these
phenomena invokes GR effects in the inner accretion disk. It has also
been proposed that the high frequency QPOs may be caused by an
inertial-acoustic instability in the disk (\cite{che95}), which has a
non-GR origin. However, this model as proposed does not extend to
frequencies as high as the 300 Hz QPO of GRO J1655-40.

Any model for the high-frequency QPOs in microquasars must also
explain the observed amplitudes and spectral characteristics of these
QPOs and the manner in which these properties vary with the state of
the X-ray source.  In the case of GRO J1655-40, the 300 Hz QPO is
detected only in the ``very hard'' group, when the X-ray luminosity is
near $0.15 L_{Edd}$ and the power-law component dominates the
spectrum. The upper limits for 300 Hz QPOs in the other groups imply
that this QPO is at least a factor of 2-3 weaker (in \% amplitude) in
fainter states. While the power law component is the primary cause of
variations in X-ray luminosity, there is also an evolution in the
thermal component as the luminosity increases. We cannot exclude the
possibility that changes in the inner disk are actually responsible
for the 300 Hz QPO, and that the larger changes in the power law
component are an unrelated byproduct of the accretion state. In that
case, the QPO would have an amplitude $\ga$1.5\% of the disk flux,
since the deconvolution of the spectral components in terms of PCA
counting rate indicates that 33\% of the counts in the ``very hard''
state come from the disk while 67\% originate in the power law
component. The spectral constraints on the origin of the 67 Hz QPO in
GRS1915+105 are far more challenging for disk oscillation models,
since the QPO amplitude has been shown to be as high as 6\% at photon
energies above 13 keV, where the disk contributes a small fraction of
the X-ray flux (\cite{mor97}).  The multiplicity of models and the
incomplete connections between inner disk timescales and the spectral
properties of the high-frequency QPOs remain as outstanding problems
in this field.

Zhang, Cui, \& Chen (1997) have outlined GR corrections to the disk
blackbody model that allow estimation of the real temperature and
radius of the inner accretion disk from the measured values of
$T_{col}$ and the disk blackbody normalization. Assuming a distance of
3.2 kpc, our spectral results for the ``very hard'' group
(Table~\ref{tab:spec}) imply an unabsorbed flux of 3.4 $\times
10^{-8}$ erg cm$^{-2}$ \ps\ for the disk component. If we substitute
this value along with a color temperature of 1.49 keV into equation 3
of Zhang, Cui, \& Chen (1997), then the inner disk radius ($r_{in}$)
associated with the ``very hard'' group is 19 km. This value is
basically in agreement with their results (22 km), which were derived
from BATSE / ASCA observations during 1995 August (\cite{zha97}). As
pointed out by Zhang, Cui, \& Chen (1997), this small inner disk can
only be larger than the last stable orbit ($r_{last}$) for a 7 \msun\
black hole if the black hole rotation is prograde and $\ga$90\% of the
maximum value.  The RXTE spectral results for the ``very hard'' group
are consistent with this view, although some of the individual ``very
hard'' observations imply inner disk radii as small as 10 km (see
below and also \cite{tom98}), implying a maximally rotating black
hole.

On the other hand, the large {\it{increase}} in the normalization of
the disk component for the ``soft'' X-ray groups
(Table~\ref{tab:spec}) necessarily suggests $r_{in} > r_{last}$ for
the same values of black hole mass and spin. These significant
variations in $f_{DBB} T_{col}^{-4}$ for GRO J1655-40 are distinctly
different from typical X-ray novae (\cite{tan95}), and yet they
resemble the pattern of changes seen at much more rapid time scales
(i.e. tens of s) in the case of GRS1915+105 (\cite{bel97b}).  If we
consider these results literally, then we must ask why the disk should
terminate well outside the last stable orbit when the luminosity of
GRO J1655-40 ($\sim 0.05 L_{Edd}$) is dominated by the thermal
component in a radio quiet state. At present there is no
straightforward explanation; further work is being done to investigate
the spectral evolution of GRO J1655-40 at shorter time scales
(\cite{sob98}). These results demonstrate the need to monitor the
spectral evolution of black hole binaries in order to scrutinize the
inner disk region. The disk blackbody model with GR corrections may or
may not provide a quantitative and reliable estimate of the inner disk
radius. Nevertheless, it does appear that the soft X-ray components
in the microquasars GRO J1655-40 and GRS1915+105 vary much more in
temperature and normalization than do the soft components in typical
X-ray novae.

While our analysis of QPO properties in GRO J1655-40 motivated a
segregation of RXTE observations into groups defined by PCA hardness
ratio, there is, in fact, a good correlation between these groups and
the canonical states of black hole binaries (\cite{van94};
\cite{van95}). The ``soft'' groups correspond with the ``high state''
seen at 1-20 keV, while ``hard'' and ``very hard'' groups display the
QPOs and the strong power law component (with index $\sim 2.5$) that
define the ``very high state''. Previous observations of the very high
state have been limited to transient conditions in GX339-4 and GS
1124-68 (X-ray Nova Muscae 1991). There is also a suggestion that the
chronic QPOs and high luminosity in GRS1915+105 constitute some kind
of variations of the very high state (\cite{mor97}). The QPO
characteristics of GRO J1655-40 appear to bridge the phenomenology of
GS 1124-68 and GRS1915+105, and these RXTE results provide a large
increase in the historical archive for the canonical form of the very
high state.

Following up on the similarity of particular RXTE observations of GRO
J1655-40 and GRS1915+105 presented in Section 3.4 and
Figure~\ref{fig:lcflick}, we modeled the average spectrum for each
observation as described in Section 4. We then computed the
luminosity, assuming distances of 3.2 and 12.5 kpc, respectively. For
GRO J1655-40, we find $T_{col}$ = 1.71 keV, the disk blackbody
normalization is 172, the photon index is 2.50, the unabsorbed total
luminosity is 2.0 $\times 10^{38}$ erg \ps\ at 1--25 keV, or $L_X \sim
0.22 L_{Edd}$, and the corrected value for the inner disk radius is
$r_{in} \sim 10$ km. In the case of GRS1915+105, we find $T_{col}$ =
1.70 keV, the disk blackbody normalization is 338, the photon index is
3.17, the unabsorbed $L_X \sim 2.9 \times 10^{39}$ erg \ps\ at 1--25
keV, and $r_{in} \sim 55$ km (assuming maximum rotation and a disk
inclination of 70$^\circ$). We may speculate that the conditions
in these two microquasars are similar with respect to either the inner
disk, with $r_{in} = r_{last}$ for maximally rotating black holes, or
with respect to the value of $L_{Edd}$. In such cases, GRS1915+105
contains the more massive black hole, with $M_1$ in the range of
38--70 \msun.

We have shown that the full composition of QPOs in GRO J1655-40
includes another quasi-stable frequency in the form of a broad QPO
(1--3\% amplitude) at 9 Hz. There is no obvious scheme in which to
asign the system's natural time scales to both the 9 and 300 Hz
oscillations for the same values of black hole mass and spin
rate. Moreover, the presence of the 9 Hz QPO complicates the
application of the frame-dragging model to QPOs in black hole binaries
(\cite{cui98}). Here, an 8 Hz QPO in GS 1124-68 was interpreted as
being analogous to the 300 Hz QPO in GRO J1655-40.  However, in this
paper we have shown that there are two low-frequency QPOs in GRO
J1655-40 that are much closer in frequency to the 8 Hz QPO in GS
1124-68, confusing the selection of QPOs that might represent the
frame dragging effect. Moreover, of the four QPO tracks in GRO
J1655-40, only the 9 Hz QPO appears to have the ``soft'' X-ray
spectrum expected for a purely disk-rooted oscillation. Clearly there
is a need for additional opportunities to observe X-ray sources in the
very high state, as we continue to make progress in learning how to
use X-ray QPOs to investigate the properties of black holes and also
to determine the origin of the hard X-ray photons in black hole
binaries.

\acknowledgments 
This work was supported, in part, by NASA grant NAG5-3680. 
Partial support for J.E.M. was provided by the Smithsonian
Institution Scholarly Studies Program. C.B. acknowledges 
support from an NSF National Young Investigator award.

\clearpage

\begin{center}
{\bf Figure Legends}
\end{center}

\figcaption[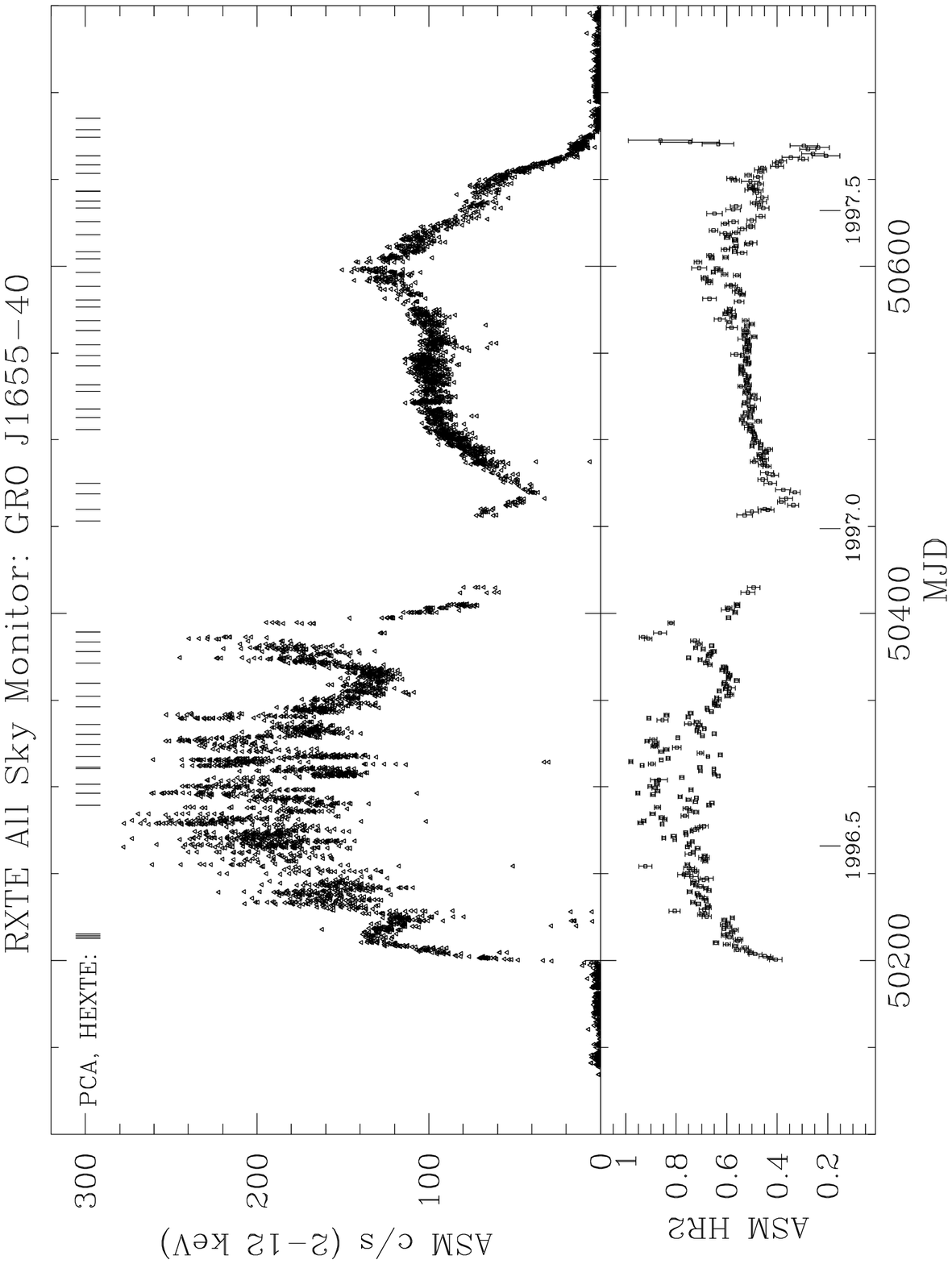]{RXTE ASM light curves and hardness ratio covering
the 1996-1997 X-ray outbrust of GRO J1655-40. For reference, the
count rate for the Crab Nebula is 75.5 ASM c/s. The ASM HR2 value is
the ratio of source counts at 5--12 keV relative to the counts at 3--5
keV. The tick marks in the upper portion of the top panel indicate the
times of RXTE pointed observations considered in this
paper. \label{fig:asm}}

\figcaption[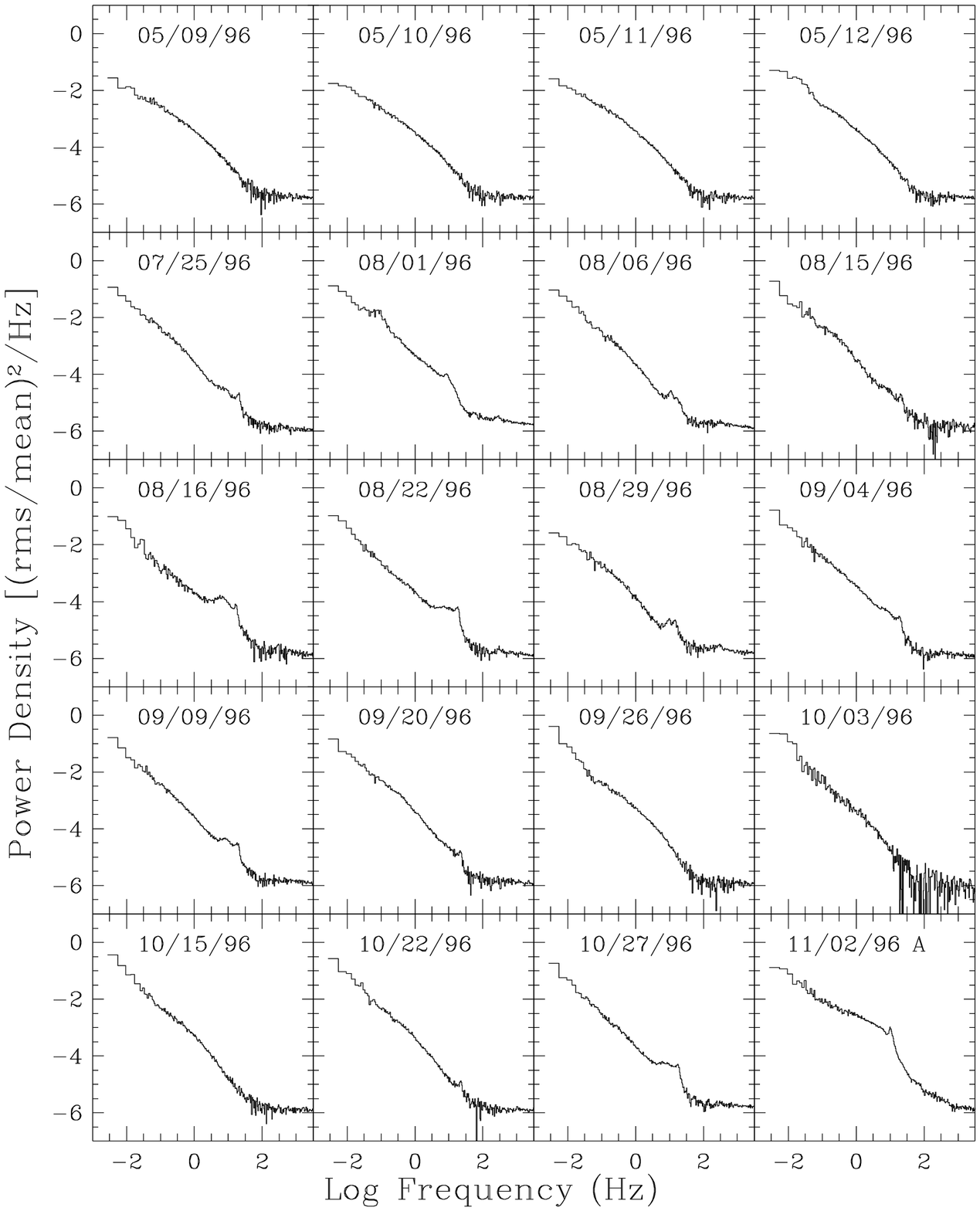]{PCA power spectra of GRO J1655-40 for the first
20 observations listed in Tables 1 and 2.  The Poisson noise level,
corrected for instrument dead time, has been subtracted. \label{fig:pds1}}

\figcaption[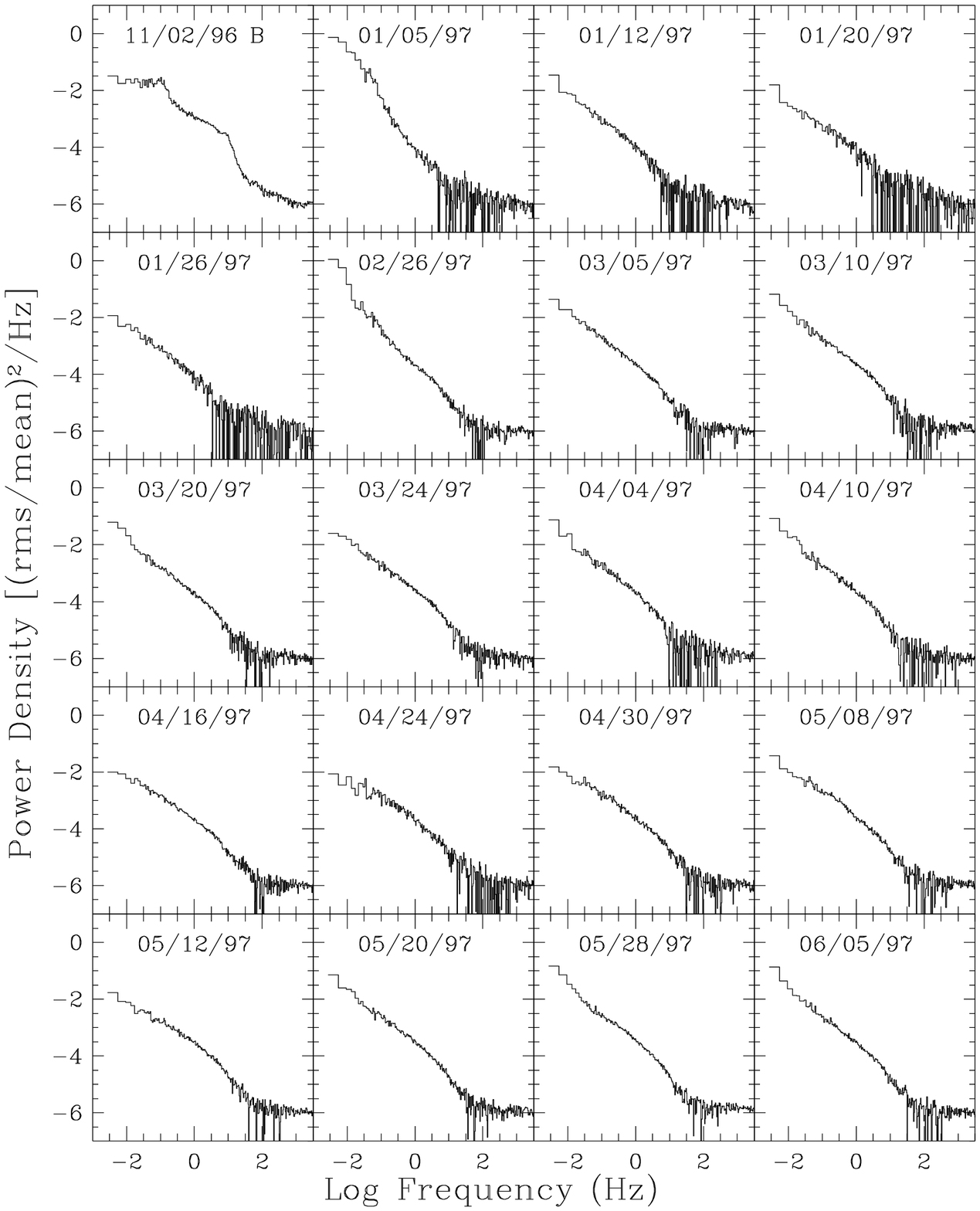]{Same as Fig. 2, but for the
subsequent 20 observations. \label{fig:pds2}}

\figcaption[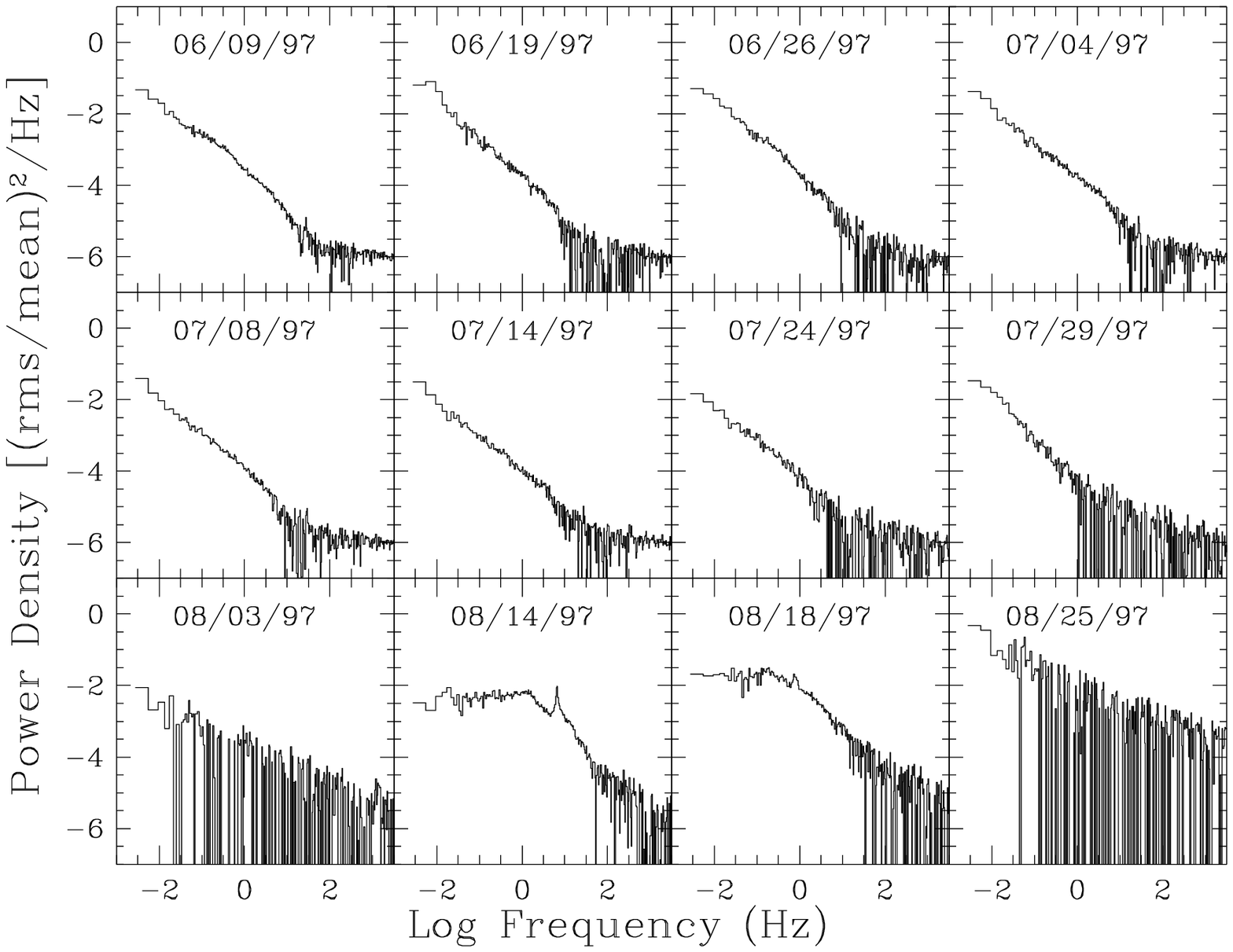]{Same as Fig. 2, but for the
last 12 RXTE observations. \label{fig:pds3}}

\figcaption[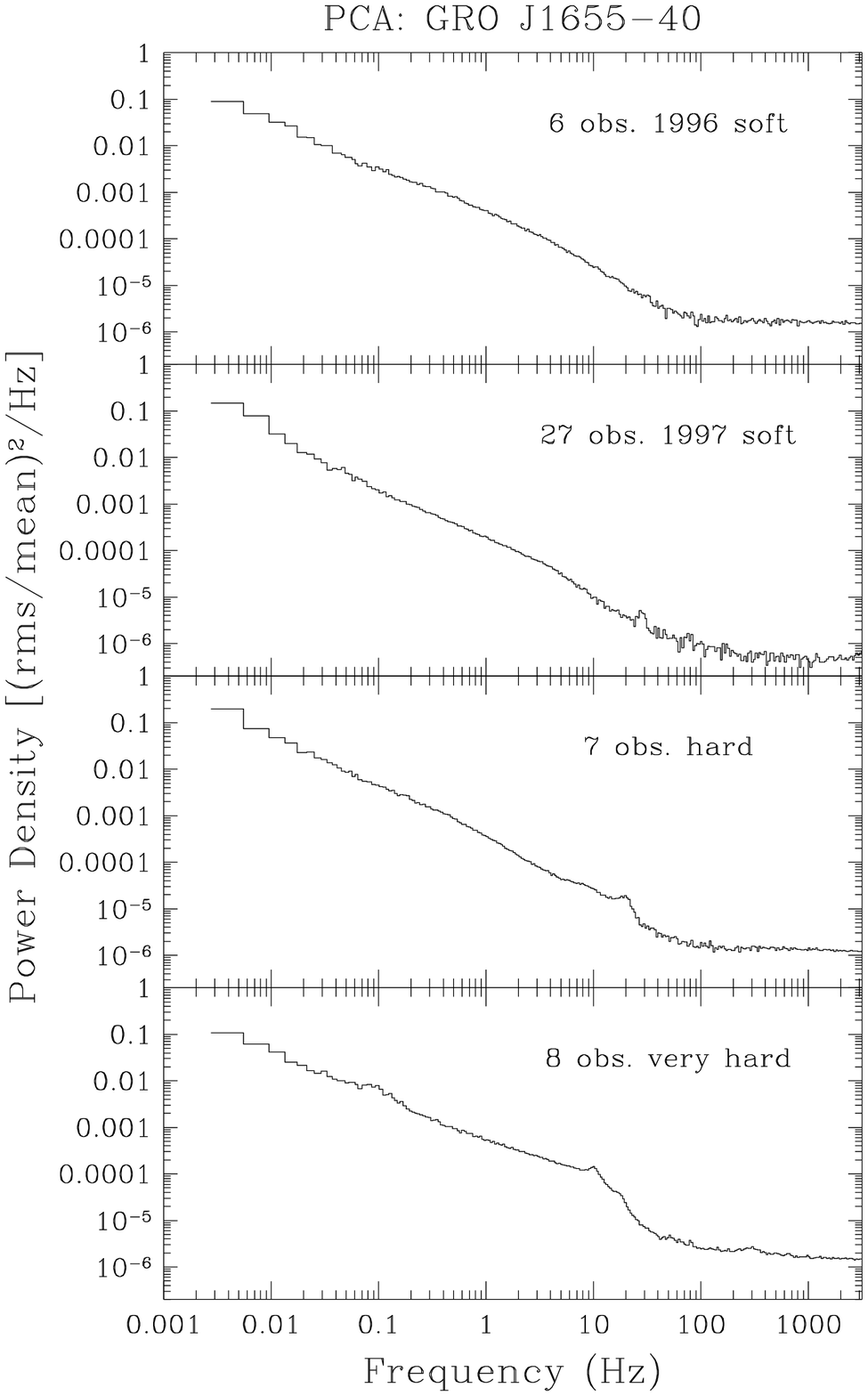]{Combined power spectra of GRO J1655-40 in
which the first 49 PCA observations are combined into 4 groups
organized by intervals of PCA HR1 and the year of observations (see
text). \label{fig:pdssum}}

\figcaption[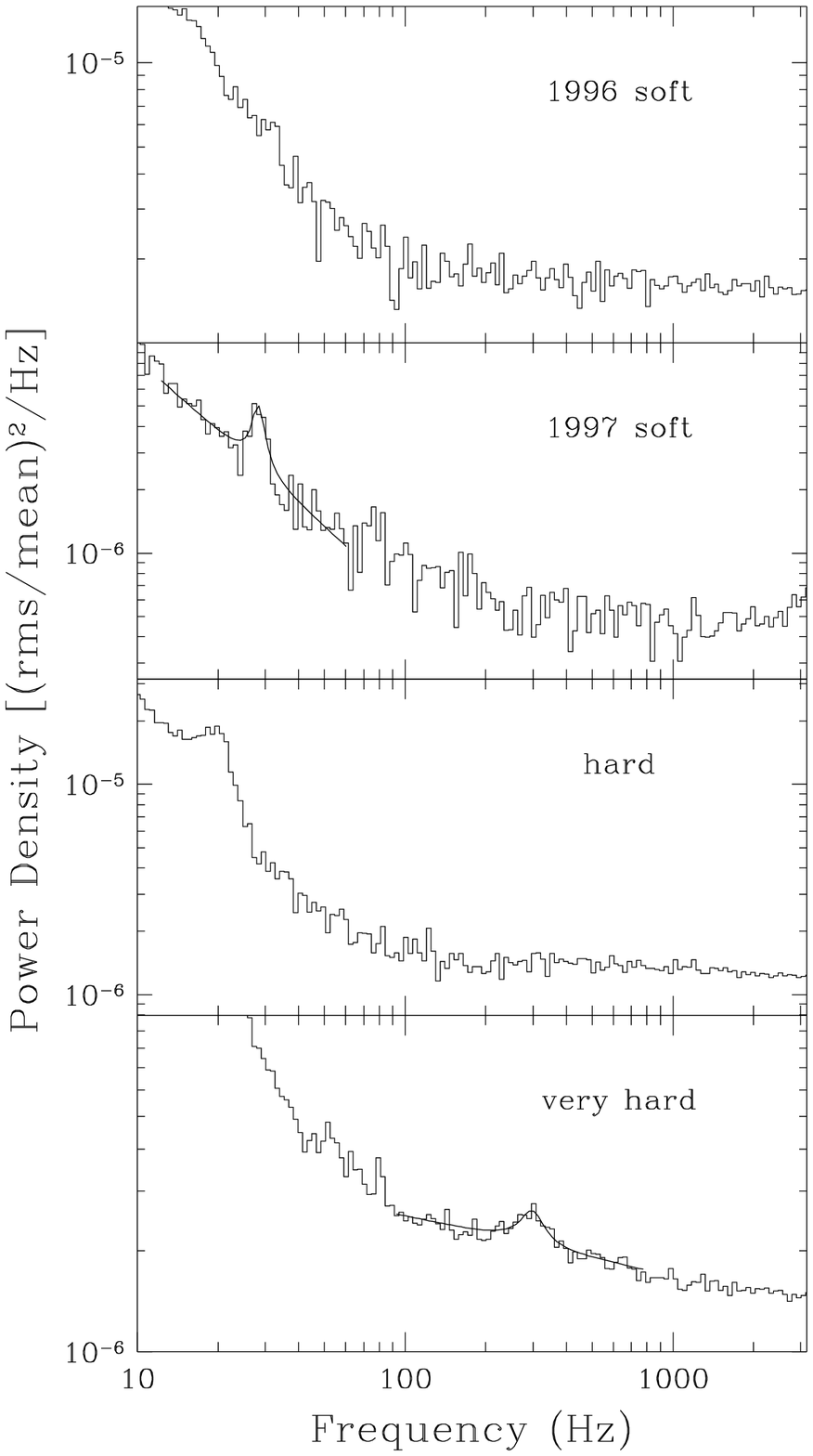]{Magnified view of a portion of Fig. 5. A QPO 
appears at 300 Hz during the 8 observations with the hardest X-ray spectra. 
\label{fig:pdssumz}}

\figcaption[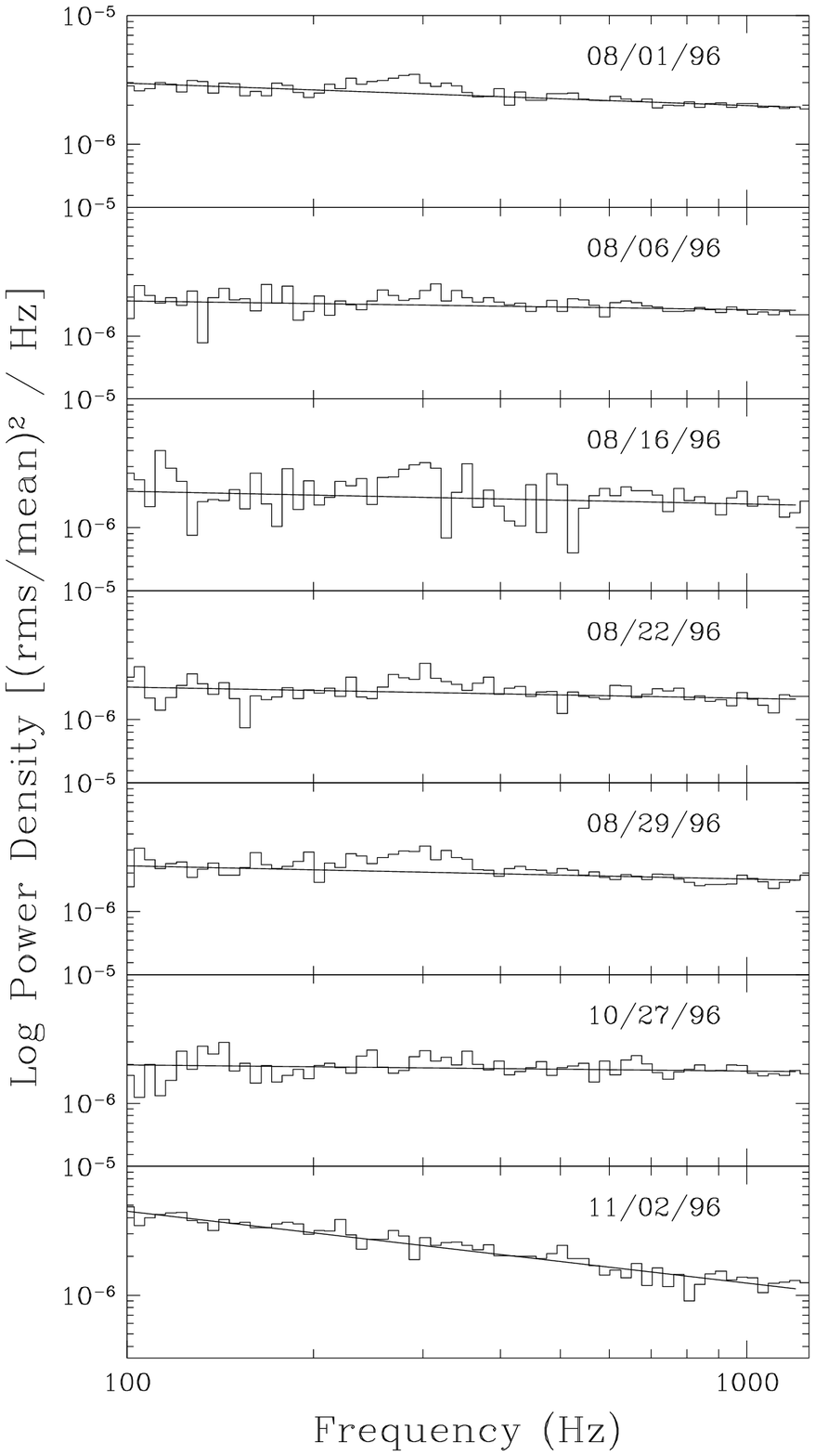]{The power spectra in the ``very hard''
group are shown individually (histograms) along with the fit to the
local power continuum (solid line). In this figure, the A and B
observations of 11/02/96 have been combined. The 300 Hz QPO appears
weakly visible in the upper 5 or 6 panels, as would be expected from
statistical effects when observing a stationary process.
\label{fig:pdsstack}}

\figcaption[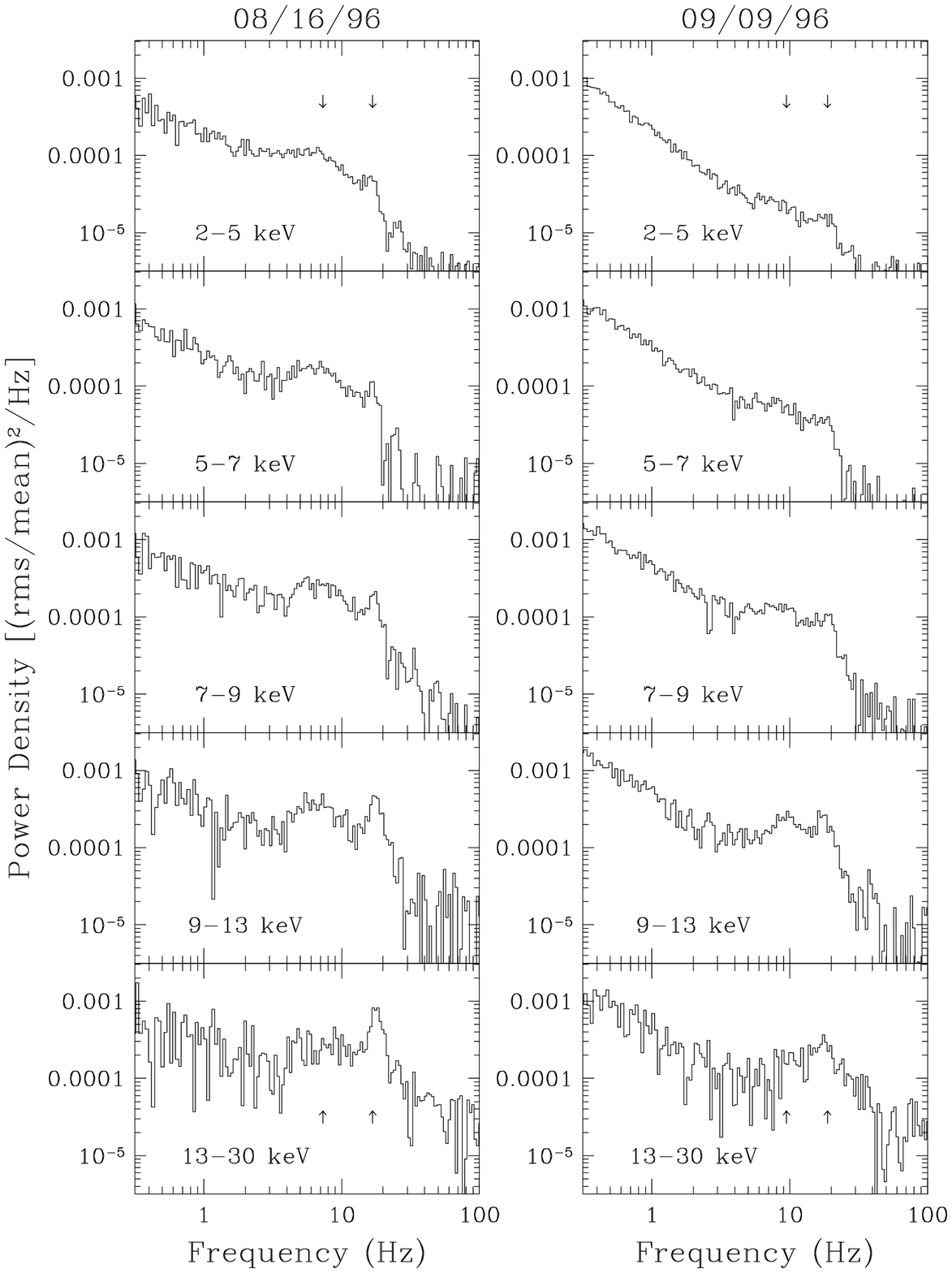]{Energy-resolved power spectra of two
observations among those that appear to have both broad and narrow
QPOs. The left panels show results for an observation (\#9) in the
``very hard'' group, while the right panels show results for an
observation (\#13) in the ``hard'' group. There appears to be a broad
and relatively soft QPO near 8--10 Hz combined with a more narrow and
hard QPO at higher frequency. The central frequencies for the 2-QPO
fits for each observation are shown with arrows in the top and bottom
panels. \label{fig:pdsen}}

\figcaption[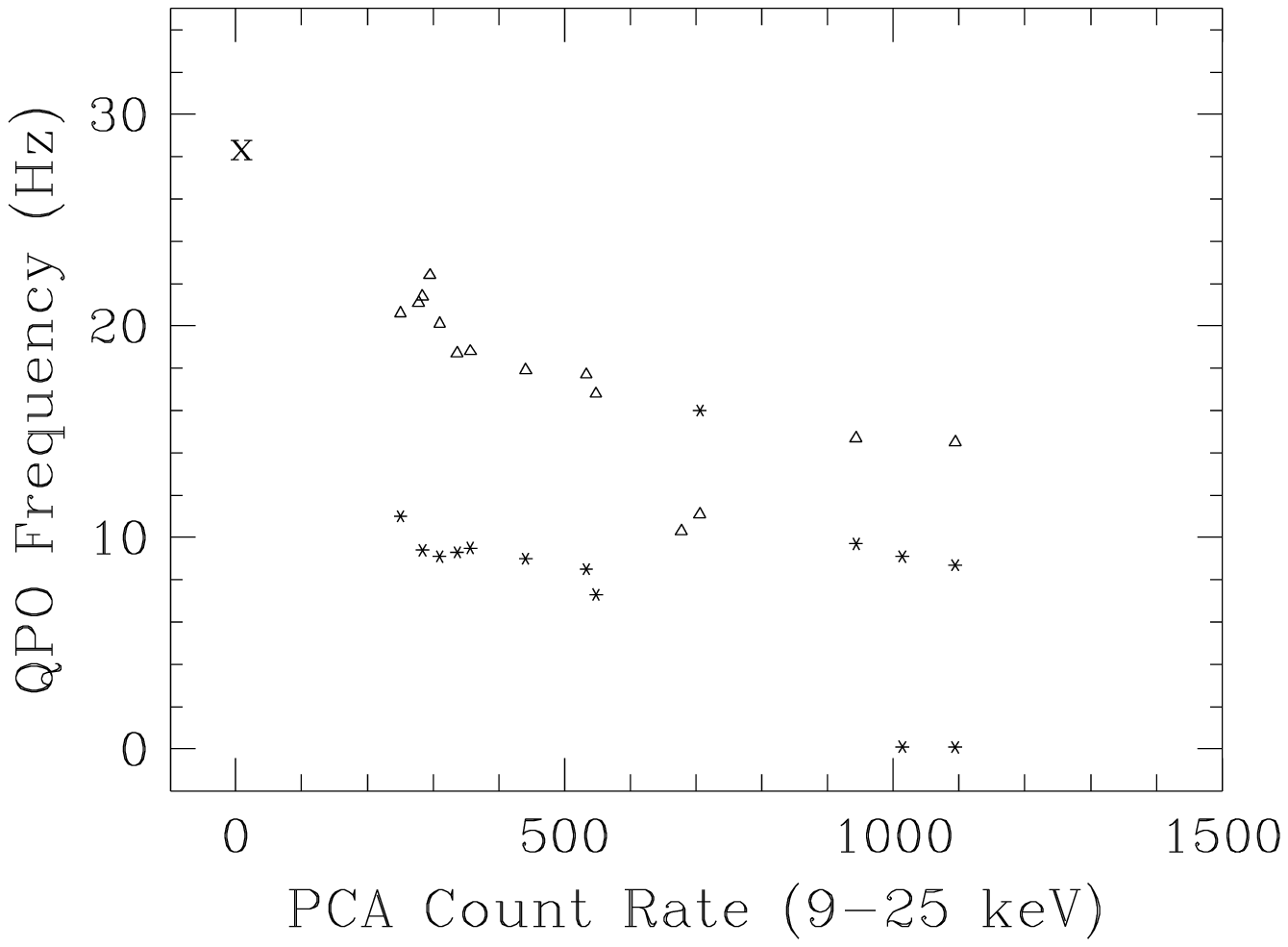]{QPO frequency vs. the PCA count rate in the
hard band (9--25 keV). The results are determined from 2-QPO fits to
the power spectra for the range of 6--35 Hz.  The ``*'' symbol is used
for broad QPOs, in which $Q < 3.0$, while the open triangles denote
narrow QPOs with $Q > 3.0$ (see text). The data point plotted with ``x''
represents the weak QPO and the mean hard count rate from the entire ``soft
1997'' group (see Figure 6). \label{fig:fxqpo}}

\figcaption[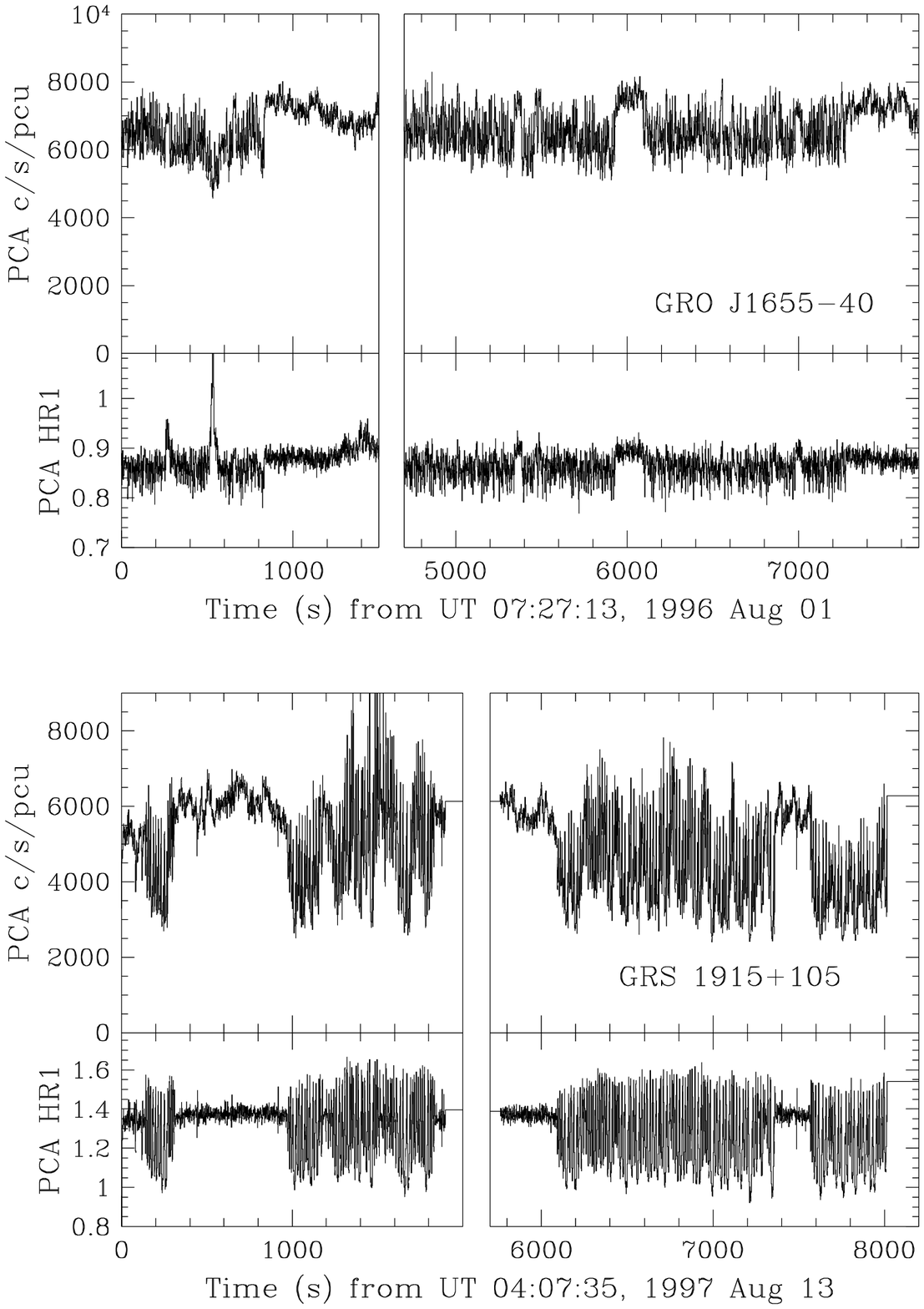] {PCA observation of GRO J1655-40 on 1996 August
1 (top panels) showing the light curve (2--25 keV) and the values of
the hardness ratio, PCA HR1. Very similar results were obtained during
a PCA observation of GRS 1915+105 on 1997 August 13 (lower panels).
In both cases, the X-ray emission teeters between a relatively steady
emission state and another state with rapid oscillations at 0.1
Hz. \label{fig:lcflick}}

 
\clearpage
 \begin{deluxetable}{rcccccc}
\footnotesize
\tablecaption{Observations of GRO J1655-40 with RXTE \label{tab:obs}}
\tablewidth{0pt}
\tablehead{
\colhead{Obs.} & \colhead{MJD start} & \colhead{MJD end} & \colhead{UT Date} &
\colhead{\#PCUs} & \colhead{exposure} & \colhead{modes\tablenotemark{a}}
}
\startdata
1  & 50212.670 & 50212.951 & 05/09/96  & 3.0 & 10122 & a \\
2  & 50213.385 & 50213.618 & 05/10/96  & 3.0 & 11429 & a \\
3  & 50214.318 & 50214.685 & 05/11/96  & 3.0 & 14039 & b \\
4  & 50215.279 & 50215.618 & 05/12/96  & 3.0 & 14016 & b \\
5  & 50289.355 & 50289.491 & 07/25/96  & 5.0 &  7430 & b \\
6  & 50296.311 & 50296.506 & 08/01/96  & 4.0 &  8903 & b \\
7  & 50301.664 & 50301.789 & 08/06/96  & 5.0 &  6335 & b \\
8  & 50310.599 & 50310.678 & 08/15/96  & 4.0 &  3415 & b \\
9  & 50311.390 & 50311.412 & 08/16/96  & 5.0 &  1895 & b \\
10 & 50317.441 & 50317.548 & 08/22/96  & 5.0 &  6153 & b \\
11 & 50324.379 & 50324.473 & 08/29/96  & 4.0 &  4927 & b \\
12 & 50330.253 & 50330.353 & 09/04/96  & 5.0 &  6189 & b \\
13 & 50335.912 & 50336.026 & 09/09/96  & 5.0 &  8164 & b \\
14 & 50346.194 & 50346.299 & 09/20/96  & 5.0 &  6438 & b \\
15 & 50352.197 & 50352.333 & 09/26/96  & 5.0 &  6599 & b \\
16 & 50359.563 & 50359.741 & 10/03/96  & 4.6 &  5426 & b \\
17 & 50371.421 & 50371.579 & 10/15/96  & 4.7 &  5385 & b \\
18 & 50378.081 & 50378.213 & 10/22/96  & 5.0 &  6887 & b \\
19 & 50383.564 & 50383.717 & 10/27/96  & 4.0 &  6265 & b \\
20 & 50389.217 & 50389.257 & 11/02/96A & 5.0 &  3457 & b \\
21 & 50389.288 & 50389.324 & 11/02/96B & 5.0 &  3015 & b \\
22 & 50453.331 & 50453.428 & 01/05/97  & 5.0 &  6551 & c \\
23 & 50460.002 & 50460.130 & 01/12/97  & 5.0 &  6536 & c \\
24 & 50468.994 & 50469.101 & 01/20/97  & 5.0 &  6912 & c \\
25 & 50474.822 & 50474.951 & 01/26/97  & 5.0 &  5885 & c \\
26 & 50505.814 & 50505.979 & 02/26/97  & 5.0 &  9638 & c \\
27 & 50512.758 & 50512.858 & 03/05/97  & 5.0 &  5884 & c \\
28 & 50517.688 & 50517.790 & 03/10/97  & 4.0 &  5831 & c \\
\tablebreak
29 & 50527.858 & 50527.998 & 03/20/97  & 5.0 &  6299 & c \\
30 & 50531.703 & 50531.799 & 03/24/97  & 5.0 &  6001 & c \\
31 & 50542.650 & 50542.715 & 04/04/97  & 4.0 &  3157 & c \\
32 & 50548.520 & 50548.577 & 04/10/97  & 5.0 &  2841 & c \\
33 & 50554.770 & 50554.917 & 04/16/97  & 5.0 &  8577 & c \\
34 & 50562.797 & 50562.821 & 04/24/97  & 5.0 &  1536 & c \\
35 & 50568.581 & 50568.657 & 04/30/97  & 5.0 &  3642 & c \\
36 & 50576.462 & 50576.549 & 05/08/97  & 4.3 &  4092 & c \\
37 & 50580.650 & 50580.738 & 05/12/97  & 5.0 &  5292 & c \\
38 & 50588.353 & 50588.458 & 05/20/97  & 5.0 &  4594 & c \\
39 & 50596.319 & 50596.424 & 05/28/97  & 4.0 &  6919 & c \\
40 & 50604.268 & 50604.363 & 06/05/97  & 5.0 &  5647 & c \\
41 & 50608.401 & 50608.572 & 06/09/97  & 5.0 &  5965 & c \\
42 & 50618.339 & 50618.365 & 06/19/97  & 5.0 &  2187 & c \\
43 & 50625.825 & 50625.879 & 06/26/97  & 5.0 &  2392 & c \\
44 & 50633.482 & 50633.568 & 07/04/97  & 5.0 &  5129 & c \\
45 & 50637.480 & 50637.649 & 07/08/97  & 5.0 &  8993 & c \\
46 & 50643.211 & 50643.489 & 07/14/97  & 5.0 &  7204 & c \\
47 & 50653.504 & 50653.638 & 07/24/97  & 5.0 &  6047 & c \\
48 & 50658.351 & 50658.422 & 07/29/97  & 5.0 &  3918 & c \\
49 & 50663.664 & 50663.672 & 08/03/97  & 3.0 &   630 & c \\
50 & 50674.438 & 50674.473 & 08/14/97  & 5.0 &  2973 & c \\
51 & 50678.566 & 50678.607 & 08/18/97  & 5.0 &  3511 & c \\
52 & 50685.434 & 50685.475 & 08/25/97  & 4.0 &  3504 & c \\
\enddata
\tablenotetext{a}{The three PCA EA modes used for GRO J1655-40 are
defined as follows: a (B\_4MS\_8A\_0\_49\_H E\_1US\_4A\_50\_8S), b
(B\_2MS\_4B\_0\_35\_H E\_16US\_16B\_36\_1S SB\_62US\_0\_35\_500MS),
and c (SB\_125US\_0\_13\_1S SB\_125US\_14\_23\_1S
SB\_125US\_24\_35\_1S E\_16US\_16B\_36\_1S).}

\end{deluxetable}

\clearpage

\begin{deluxetable}{rcccccccl}
\footnotesize
\tablecaption{PCA Count Rates and QPO Detections \label{tab:results}}
\tablewidth{0pt}
\tablehead{
\colhead{Obs.} & \colhead{UT Date} & \colhead{PCA} & \colhead{$\sigma$} &
\colhead{PCA} & \colhead{PCA} & \colhead{Soft} & \colhead{Hard} & \colhead{QPOs and} \\
\colhead{} & \colhead{} & \colhead{rate\tablenotemark{a}} & \colhead{1 s bins} & \colhead{HR1\tablenotemark{b}} & \colhead{HR2\tablenotemark{c}} & \colhead{2-5 keV} & \colhead{9-25 keV} & \colhead{Comments}
}
\startdata
1  & 05/09/96 & 3556 & 156.8 & 0.417 & 0.075 & 2473 &  84 & - \\
2  & 05/10/96 & 3724 & 130.8 & 0.413 & 0.075 & 2540 &  85 & - \\
3  & 05/11/96 & 3649 & 133.5 & 0.416 & 0.075 & 2502 &  85 & - \\
4  & 05/12/96 & 3671 & 157.2 & 0.422 & 0.076 & 2490 &  88 & - \\
5  & 07/25/96 & 4721 & 289.6 & 0.468 & 0.155 & 2971 & 295 & 9.1,20.1 \\
6  & 08/01/96 & 7139 & 973.7 & 0.680 & 0.284 & 3511 &1006 & 0.09,8.7,14.5,281.1 \\
7  & 08/06/96 & 6013 & 328.4 & 0.594 & 0.236 & 3307 & 657 & 11.1,16.0,313.5 \\
8  & 08/15/96 & 4327 & 343.8 & 0.476 & 0.157 & 2719 & 280 & 9.4, 21.4 \\
9  & 08/16/96 & 5165 & 252.4 & 0.524 & 0.224 & 3004 & 514 & 7.3,16.8,292.5 \\
10 & 08/22/96 & 4937 & 305.5 & 0.501 & 0.198 & 2977 & 417 & 9.0,17.9,301.8 \\
11 & 08/29/96 & 6820 & 246.1 & 0.639 & 0.265 & 3547 & 875 & 9.7,14.7,295.4 \\
12 & 09/04/96 & 4727 & 337.5 & 0.471 & 0.167 & 2948 & 322 & 9.3,18.7 \\
13 & 09/09/96 & 4734 & 313.8 & 0.483 & 0.174 & 2921 & 341 & 9.5,18.8 \\
14 & 09/20/96 & 4119 & 245.0 & 0.452 & 0.147 & 2632 & 241 & 20.6 \\
15 & 09/26/96 & 3854 & 293.9 & 0.432 & 0.106 & 2554 & 149 & - \\
16 & 10/03/96 & 3806 & 281.6 & 0.447 & 0.121 & 2521 & 180 & - \\
17 & 10/15/96 & 4595 & 383.3 & 0.481 & 0.142 & 2884 & 266 & 21.1 \\
18 & 10/22/96 & 4766 & 412.9 & 0.469 & 0.147 & 3026 & 282 & 22.4 \\
19 & 10/27/96 & 5775 & 411.2 & 0.531 & 0.197 & 3377 & 497 & 8.5,17.7,316.7 \\
20 & 11/02/96A& 4328 & 354.5 & 0.609 & 0.299 & 2257 & 632 & 10.3 \\
21 & 11/02/96B& 6350 & 457.0 & 0.658 & 0.294 & 3209 & 935 & 0.10,9.1 \\
22 & 01/05/97 & 1754 & 183.3 & 0.352 & 0.127 & 1220 &  82 & -; dips \\
23 & 01/12/97 & 1589 &  47.6 & 0.308 & 0.093 & 1160 &  48 & - \\
24 & 01/20/97 &  890 &  21.6 & 0.228 & 0.040 &  708 &   8 & - \\
25 & 01/26/97 & 1123 &  28.3 & 0.259 & 0.070 &  856 &  22 & - \\
26 & 02/26/97 & 2263 & 239.3 & 0.333 & 0.049 & 1650 &  30 & -; dips \\
27 & 03/05/97 & 2533 &  89.0 & 0.343 & 0.056 & 1831 &  40 & - \\
28 & 03/10/97 & 2574 &  97.5 & 0.367 & 0.063 & 1836 &  50 & - \\
\tablebreak
29 & 03/20/97 & 2671 & 107.0 & 0.359 & 0.071 & 1863 &  60 & - \\
30 & 03/24/97 & 2398 &  84.3 & 0.340 & 0.051 & 1736 &  33 & - \\
31 & 04/04/97 & 2671 & 105.7 & 0.356 & 0.071 & 1891 &  59 & - \\
32 & 04/10/97 & 2715 & 107.1 & 0.361 & 0.075 & 1915 &  65 & - \\
33 & 04/16/97 & 2538 &  67.3 & 0.345 & 0.053 & 1833 &  36 & - \\
34 & 04/24/97 & 2583 &  68.5 & 0.366 & 0.056 & 1860 &  41 & - \\
35 & 04/30/97 & 2649 &  78.3 & 0.356 & 0.055 & 1898 &  40 & - \\
36 & 05/08/97 & 2874 &  89.1 & 0.389 & 0.062 & 2022 &  53 & - \\
37 & 05/12/97 & 2884 &  89.4 & 0.372 & 0.061 & 2022 &  49 & - \\
38 & 05/20/97 & 3154 & 128.1 & 0.386 & 0.068 & 2205 &  65 & - \\
39 & 05/28/97 & 3342 & 174.4 & 0.416 & 0.084 & 2217 &  94 & - \\
40 & 06/05/97 & 3030 & 146.1 & 0.377 & 0.069 & 2128 &  63 & - \\
41 & 06/09/97 & 2872 & 105.0 & 0.366 & 0.062 & 1996 &  51 & - \\
42 & 06/19/97 & 2599 & 154.2 & 0.360 & 0.075 & 1823 &  63 & - \\
43 & 06/26/97 & 2099 & 174.4 & 0.321 & 0.048 & 1584 &  27 & - \\
44 & 07/04/97 & 1824 &  65.1 & 0.302 & 0.051 & 1356 &  25 & - \\
45 & 07/08/97 & 1818 &  62.7 & 0.317 & 0.075 & 1324 &  43 & - \\
46 & 07/14/97 & 1674 &  55.4 & 0.289 & 0.052 & 1256 &  23 & - \\
47 & 07/24/97 & 1149 &  29.3 & 0.268 & 0.087 &  864 &  30 & - \\
48 & 07/29/97 &  811 &  25.6 & 0.253 & 0.121 &  609 &  31 & - \\
49 & 08/03/97 &  463 &  14.2 & 0.210 & 0.109 &  358 &  13 & - \\
50 & 08/14/97 &  213 &  13.5 & 0.748 & 0.392 &   95 &  47 & 1.4,6.4 \\
51 & 08/18/97 &   61 &   8.4 & 0.945 & 0.475 &   23 &  18 & 0.2,0.8 \\
52 & 08/25/97 &  3.5 &    -  & 0.693 & 0.411 &  1.8 & 0.5 & - \\
\enddata 

\tablenotetext{a}{PCA source rate at full bandwidth (effectively 2--25 keV) 
in cts s$^{-1}$ pcu$^{-1}$.}
\tablenotetext{b}{Ratio of 5--9 keV and 2--5 keV source count rates.}
\tablenotetext{c}{Ratio of 9--13 keV and 5--9 keV count rates.}

\end{deluxetable}

\clearpage

\begin{deluxetable}{cccccc}
\footnotesize
\tablecaption{QPO fit parameters \label{tab:qpos}}
\tablewidth{0pt}
\tablehead{
\colhead{Obs.} & \colhead{Date} & \colhead{$\nu$} & \colhead{FWHM} &
\colhead{Q} & \colhead{amplitude}
}
\startdata
5  & 07/25/96  &  9.1  & 5.4   & 1.7 & 0.0095 \\
5  & 07/25/96  & 20.1  & 5.4   & 3.7 & 0.0110 \\
6  & 08/01/96  & 0.086 & 0.058 & 1.5 & 0.0287 \\
6  & 08/01/96  &  8.7  & 4.1   & 2.1 & 0.0189 \\
6  & 08/01/96  & 14.5  & 4.5   & 3.2 & 0.0090 \\
7  & 08/06/96  & 11.1  & 1.5   & 7.3 & 0.0054 \\
7  & 08/06/96  & 16.0  & 10.5  & 1.5 & 0.0123 \\
8  & 08/15/96  &  9.4  & 5.0   & 1.9 & 0.0084 \\
8  & 08/15/96  & 21.4  & 4.7   & 4.6 & 0.0078 \\
9  & 08/16/96  &  7.3  & 7.2   & 1.0 & 0.0332 \\
9  & 08/16/96  & 16.8  & 2.5   & 6.7 & 0.0132 \\
10 & 08/22/96  &  9.0  & 8.3   & 1.1 & 0.0225 \\
10 & 08/22/96  & 17.9  & 3.5   & 5.1 & 0.0173 \\
11 & 08/29/96  &  9.7  & 3.4   & 2.9 & 0.0081 \\
11 & 08/29/96  & 14.7  & 3.1   & 4.7 & 0.0078 \\
12 & 09/04/96  &  9.3  & 7.0   & 1.3 & 0.0146 \\
12 & 09/04/96  & 18.7  & 4.5   & 4.2 & 0.0127 \\
13 & 09/09/96  &  9.5  & 7.4   & 1.3 & 0.0154 \\
13 & 09/09/96  & 18.8  & 4.7   & 4.0 & 0.0110 \\
14 & 09/20/96  & 20.6  & 3.4   & 6.1 & 0.0076 \\
17 & 10/15/96  & 21.1  & 2.2   & 9.6 & 0.0032 \\
18 & 10/22/96  & 22.4  & 3.3   & 6.8 & 0.0058 \\
19 & 10/27/96  &  8.5  & 7.0   & 1.2 & 0.0180 \\
19 & 10/27/96  & 17.7  & 4.5   & 3.9 & 0.0151 \\
20 & 11/02/96A & 10.3  & 2.2   & 4.7 & 0.0454 \\
21 & 11/02/96B & 0.100 & 0.082 & 1.2 & 0.0442 \\
21 & 11/02/96B &  9.1  & 4.4   & 2.1 & 0.0279 \\
22-48  &  soft 97  & 28.3  & 3.6   & 7.9 & 0.0031 \\
\enddata 
\end{deluxetable}

\clearpage

\begin{deluxetable}{lcrrccccrccr}
\footnotesize
\tablecaption{RXTE Spectral Parameters for GRO J1655-40 \label{tab:spec}}
\tablewidth{0pt}
\tablehead{
\colhead{Group} & \colhead{\#} & \colhead{time} & \colhead{$\chi_{\nu}^2$} & 
\colhead{$N_H$} & \colhead{Disk $T_{col}$} & \colhead{Disk} & \colhead{P.L.} & \colhead{P.L.} & 
\colhead{$f_{DBB}$} & \colhead{$f_{PL}$}  & \colhead{$L_x$}\\
\colhead{} & \colhead{Obs.} & \colhead{(ks)} & \colhead{} & \colhead{} &
\colhead{keV} & \colhead{Norm.} & \colhead{index} & \colhead{Norm.} &
\colhead{} & \colhead{}
}
\startdata
Soft 96a  &  4 &  49.6 & 19.86 & 2.21(5) & 1.24(1) & 1570(38) & 4.50(*) & 146.0(56) & 4.43 & 0.11 &  9.2 \\
Soft 96b  &  2 &   7.6 &  1.04 & 1.12(4) & 1.26(2) & 1615(22) & 2.22(1) &  2.40(5)  & 4.24 & 0.62 &  8.3 \\
Hard      &  7 &  40.5 &  0.70 & 0.62(1) & 1.30(1) & 1220(33) & 2.46(2) & 10.61(24) & 4.23 & 1.79 & 10.2 \\
Very Hard &  8 &  40.9 &  1.27 & 1.51(3) & 1.43(1) &  418(32) & 2.67(1) & 51.9(50)  & 2.14 & 6.04 & 14.2 \\
Soft 97   & 27 & 102.2 &  4.19 & 1.88(5) & 1.10(1) & 2211(31) & 2.42(3) &  0.64(7)  & 3.00 & 0.10 &  5.4 \\
\\
Soft 96a  &  4 &  49.6 & 24.01 & 0.90 fix & 1.26(1) & 1392(29) & 4.24(2) & 64.9(9)  & 3.85 & 0.83 &  9.4 \\
Soft 96b  &  2 &   7.6 &  1.14 & 0.90 fix & 1.26(2) & 1570(24) & 2.20(2) &  2.23(5) & 4.31 & 0.59 &  8.4 \\
Hard      &  8 &  40.5 &  0.84 & 0.90 fix & 1.29(1) & 1366(28) & 2.49(1) & 11.7(3)  & 4.16 & 1.81 & 10.1 \\
Very Hard &  7 &  40.9 &  1.77 & 0.90 fix & 1.49(2) &  376(13) & 2.62(2) & 47.5(9)  & 2.41 & 5.89 & 14.3 \\
Soft 97   & 27 & 102.2 &  6.71 & 0.90 fix & 1.12(1) & 1950(21) & 2.30(3) &  0.47(4) & 3.05 & 0.10 &  5.6 \\
\enddata  
\tablecomments{The fits with $N_H$ as a free parameter have 49 degrees of freedom. 
 The uncertainties given in parentheses are in units of the most significant digit.
 The '*' in row 1 indicates an upper limit imposed to combat an unstable fit. 
 The units for N$_H$ (col. 5) are  10$^{22}$ cm$^{-2}$.  Flux units (cols. 10, 11) 
 are $10^{-8}$ erg cm$^{-2}$ \ps at 2--25 keV. Col. 12 lists the unabsorbed luminosity 
 in units of $10^{37}$ erg \ps at 1.3--25 keV, for $d=3.2$ kpc. }

\end{deluxetable}

\clearpage
\plotone{asm16.ps}
\clearpage
\plotone{pds1.ps}
\clearpage
\plotone{pds2.ps}
\clearpage
\plotone{pds3.ps}
\clearpage
\plotone{p1_sumfft1.ps}
\clearpage
\plotone{p1_sumfft2.ps}
\clearpage
\plotone{p1_stack300.ps}
\clearpage
\plotone{p1_pdsen.ps}
\clearpage
\plotone{pub_fxqpo.ps}
\clearpage
\plotone{pub_lc.ps}

\end{document}